# $SU(2)$ ABELIAN PROJECTED ACTION FOR RENORMALISATION GROUP FLOWS

FILIPE FREIRE
*School of Theoretical Physics, Dublin Inst. for Adv. Studies*
*and*
*Dept. of Mathematical Physics, NUI Maynooth, Ireland*

**Abstract.** The dual Meissner effect scenario of confinement is analysed using exact renormalisation group (ERG) equations. In particular, the low energy regime of $SU(2)$ Yang-Mills (YM) is studied in a maximal Abelian gauge (MAG). It is shown that under general conditions the effective action derived when integrated using ERG methods contains the relevant degrees of freedom for confinement. In addition, the physics in the confining regime is dual to that of the broken phase of an Abelian Higgs model.

**Introduction.** In the dual Meissner effect picture the vacuum of a YM gauge theory, with mass gap, consists of a magnetic condensate which leads to a linear potential that confines quarks [1]. This has been studied in the context of dual models of superconductivity where magnetic degrees of freedom are treated as particle fluctuations. However, it is unlikely that this scenario alone works because monopoles are heavy making it unlikely for a magnetic Higgs mechanism to take place. An alternative approach using the notion of Abelian projection has been developed in theories without Higgs fields [2, 3] such as QCD. This approach is taken in the present work. Abelian projection uses gauge fixing conditions to attenuate non-Abelian degrees of freedom while the maximal Abelian subgroup of the gauge group remains unbroken. The residual invariance is fixed by using Lorentz conditions for each remnant Abelian invariance.

Here a MAG condition is used for the gauge fixing. On the lattice Abelian dominance [4] has been observed in this gauge indicating the existence of an intermediate scale below which non-Abelian gauge field variables are suppressed. This separation of scales is a good feature for working in a MAG. The ERG for gauge theories [5, 6] provides an ideal framework for investigating the present scenario because of this splitting of scales at low energies. The use of ERG has already been highlighted by Ellwanger [7] who has done similar studies although using a Landau gauge, therefore not having the benefit of Abelian dominance. The ERG has been formulated for general linear gauges [8]. Though the MAG is non-linear it poses no new problems as it can be seen as a particular case of a background field gauge and therefore it is easy to transfer the results of [9].

In this talk I apply this procedure to the pure $SU(2)$ YM and derive an effective Abelian action for this theory. Then, under general conditions, it is shown, using a functional analysis based on ERG methods, that the effective action derived here contains the prerequisites for confinement. Moreover, it is also shown that the physics in the confining phase is dual to that of a type II superconductor. Technical details can be found in [10].

**The $SU(2)$ YM action.** Consider pure $SU(2)$ YM in a four dimensional Euclidean space. In conventional variables its action is $S_{\text{YM}} = \frac{1}{4} \int_x F_{\mu\nu}^a F_{\mu\nu}^a$, where $\int_x$ is a short-hand for $\int d^4 x$ and $F_{\mu\nu}^a$ is the field strength. When expressed in terms of the field variables $A_\mu = A_\mu^3$, $\phi_\mu = \frac{1}{\sqrt{2}}(A_\mu^1 - i A_\mu^2)$ and $\phi_\mu^\dagger$, respectively neutral and charged potentials, this action takes the form

$$S_{\text{YM}} = \frac{1}{4} \int_x (f_{\mu\nu} + C_{\mu\nu})(f_{\mu\nu} + C_{\mu\nu}) + \frac{1}{2} \int_x \Phi_{\mu\nu}^\dagger \Phi_{\mu\nu}. \qquad (1)$$

Here $f_{\mu\nu}$ and $\Phi_{\mu\nu}$ are the field strengths for the Abelian and the charged gauge fields, respectively, $f_{\mu\nu} = \partial_\mu A_\nu - \partial_\nu A_\mu$ and $\Phi_{\mu\nu} = D_\mu \phi_\nu - D_\nu \phi_\mu$, with $D_\mu = \partial_\mu - ig A_\mu$ the Abelian covariant derivative. Finally, the antisymmetric tensor $C_{\mu\nu}$ is a real quadratic combination of the off-diagonal fields, $C_{\mu\nu} = ig(\phi_\mu^\dagger \phi_\nu - \phi_\nu^\dagger \phi_\mu)$.

Clearly the action (1) is explicitly invariant under a $U(1)$ subgroup of $SU(2)$. The gauge is fixed by the condition, $F^\pm[\phi, A] := (\partial_\mu \pm ig A_\mu)\phi_\mu = 0$. This MAG leaves the $U(1)$ invariance in (1) unbroken. The remnant gauge freedom is fixed with the Lorentz condition, $F[A] := \partial_\mu A_\mu = 0$. Gribov copies exist for this particular choice of gauge fixing but their effects are not sizable [11] and shall be neglected at this stage.

**Abelian dominance.** The suppression of the off-diagonal fields observed on the lattice in a MAG [4] suggests that they acquire a mass dynamically. Some mass generation mechanisms have been investigated in MAGs [12, 13]. See also [14, 15]. It is expected that this mass sets a scale of Abelian dominance, say $M_{\text{AD}}$, below which a qualitative change of the relevant dynamical variables takes place. At a much lower scale this will lead to confinement. Preliminary results from the lattice give $M_{\text{AD}} \simeq 1.2$ GeV [16] which indicates that the scale of Abelian dominance is indeed larger than the confining scale $\Lambda_{\text{QCD}}$. Though the mass of the off-diagonal gauge field is gauge dependent, it does not imply that this energy scale itself is a gauge artifact. The value of $M_{AD}$ obtained in [16] is, within error margins, not far from the lightest glueball mass in $SU(2)$ YM, $m_G^{0^{++}} \sim 1.67$ GeV [17]. In $SU(3)$ YM the value of $m_G^{0^{++}}$ is in the same range, $m_G^{0^{++}} = 1.6 - 1.8$ GeV [18]. It would be interesting to look for signs of Abelian dominance at scales below $m_G^{0^{++}}$.

The first step towards an effective Abelian theory is to integrate over the off-diagonal vector fields $\phi_\mu$. Though there are vertices with four of these fields, the $C_{\mu\nu}C_{\mu\nu}$ term in (1), the integration can be performed by introducing an auxiliary antisymmetric tensor field, $B_{\mu\nu}$, via the replacement, $\frac{1}{4}\int_x C_{\mu\nu}C_{\mu\nu} \to -\frac{1}{4}\int_x \tilde{B}_{\mu\nu}\tilde{B}_{\mu\nu} + \frac{1}{2}\int_x \tilde{B}_{\mu\nu}C_{\mu\nu}$. It follows that the equation of motion for $B_{\mu\nu}$ is $\tilde{B}_{\mu\nu} = C_{\mu\nu}$. The resulting action is now quadratic in the off-diagonal fields. By fixing the MAG parameter $\xi' = 1$, these quadratic terms in $\phi_\mu$ are

$$S_{\phi^2} = \int_x \left( \phi_\mu^\dagger \, a_{\mu\nu}^{-+} \, \phi_\nu + \phi_\mu \, a_{\mu\nu}^{+-} \, \phi_\nu^\dagger + \phi_\mu \, a_{\mu\nu}^{++} \, \phi_\nu + \phi_\mu^\dagger \, a_{\mu\nu}^{--} \, \phi_\nu^\dagger \right), \qquad (2)$$

where $a_{\mu\nu}^{-+} = -\frac{1}{2}g_{\mu\nu}(D_\rho D_\rho + g^2 \, \bar{c}_+ c_+) + \frac{ig}{2}(2f_{\mu\nu} + \tilde{B}_{\mu\nu}) = [a_{\mu\nu}^{+-}]^\dagger$ and $a_{\mu\nu}^{++} = g^2 \, g_{\mu\nu} \, \bar{c}_- c_+ = [a_{\mu\nu}^{--}]^\dagger$. After a Gaussian integration the effective action receives a contribution of the form, $\frac{1}{2}\mathrm{Tr}\ln a_{\mu\nu}^{SS'}$, with $S, S' = \pm$. In a Schwinger proper-time representation using a complete basis of plane waves and performing a Taylor series expansion on the proper-time $t$, the trace takes the form [10]

$$\mathrm{Tr}\ln a_{\mu\nu}^{SS'} = -\lim_{s\to 0} \frac{d}{ds} \frac{\mu^{2s}}{\Gamma(s)} \int_0^\infty dt\, t^{s-3} \int_x \sum_{n=0}^\infty \frac{(-1)^n}{n!} \int_k e^{-\frac{1}{2}k^2} \times$$
$$\mathrm{tr}\left[ \left( t a_{\mu\nu}^{SS'} - i\sqrt{t}\, g_{\mu\nu}(\bar{\delta}^{S+}\bar{\delta}^{S'-} k_\rho D_\rho + \bar{\delta}^{S-}\bar{\delta}^{S'+} k_\rho D_\rho^\dagger) \right)^n \right]\Big|_{\mathrm{reg.}}, \qquad (3)$$

where tr denotes a trace over all indices and $\bar{\delta}^{SS'}$ is the inverse Kronecker delta, $\bar{\delta}^{SS'} = 1$ if $S \neq S'$ and $\bar{\delta}^{SS'} = 0$ if $S = S'$.

The momentum integration in (3) is convergent but not solvable. In order to proceed the Taylor series is truncated by keeping only the leading terms up to $n = 3$ inclusively in a small $t$ expansion. This provides a systematic approach for computing the leading effective vertices for short range interactions by decreasing importance in their UV relevance.

**The Abelian effective action.** On the right-hand side of (3) only the integer powers of $t$ do not vanish after the integration over the momenta, because the integrand for the half integer powers of $t$ is odd in the momenta. The $t$ integral is now splitted into two parts, $\int_0^{1/\Lambda^2} + \int_{1/\Lambda^2}^\infty$. The scale $\Lambda$ is a UV scale larger than $M_{AD}$ chosen so that the integration for $t > 1/\Lambda^2$ is suppressed by powers of $M_{AD}^2/\Lambda^2$. Therefore, in the spirit of Abelian dominance only the integration for $t < 1/\Lambda^2$ is kept. After the integration, the expansion in powers of $t$ emerges as an expansion in vertex operators of decreasing UV relevance. Here only the terms up to the first non-relevant ones, i.e. $\mathcal{O}(1/\Lambda^2)$ are kept. The resulting $U(1)$ invariant effective action, where the renormalisation scale is set $\mu = \Lambda$, is

$$\begin{aligned}
S_{\text{eff}} &= \frac{1}{4}\left(1 - \frac{5g^2\gamma}{12\pi^2}\right)\int_x f_{\mu\nu}f_{\mu\nu} - \frac{1}{4}\int_x B_{\mu\nu}\left(\frac{g^2}{96\pi^2}\frac{\Box}{\Lambda^2} + 1 + \frac{g^2\gamma}{8\pi^2}\right)B_{\mu\nu} - \\
&\quad - \frac{g^2\gamma}{8\pi^2}\int_x \tilde{f}_{\mu\nu}B_{\mu\nu} + \frac{1}{2\xi}\int_x(\partial_\mu A_\mu)^2 - \int_x \bar{c}_+\left(D_\mu^\dagger D_\mu^\dagger - \frac{g^2\Lambda^2}{2\pi^2}\right)c_+ - \\
&\quad - \int_x \bar{c}_-\left(D_\mu D_\mu - \frac{g^2\Lambda^2}{2\pi^2}\right)c_- + \frac{g^4\gamma}{2\pi^2}\int_x \bar{c}_+c_+\bar{c}_-c_- - \\
&\quad - \frac{1}{\Lambda^2}\frac{g^2}{96\pi^2}\left(\int_x \tilde{f}_{\mu\nu}\Box B_{\mu\nu} + \cdots\right) + \mathcal{O}\left(\frac{1}{\Lambda^4}, B^4\right). \quad (4)
\end{aligned}$$

The one-loop coupling and wave function corrections are the terms proportional to the Euler gamma. Note that the $\mathcal{O}(\Lambda^2)$ terms correspond to a mass renormalisation. In a more complete treatment where the hard modes of all fields are integrated out to one-loop order down to $\Lambda$, the ghost mass terms are canceled in accordance to BRS invariance.

There are two new relevant vertices in (4): the coupling between the dual of the Abelian field strength and the tensor field, and the 4-ghosts vertex. Kondo [19] has shown by using a Hodge decomposition of $B_{\mu\nu}$ that the $\tilde{f}_{\mu\nu}B_{\mu\nu}$ term encapsulates the coupling between a gauge field potential and a magnetic current $J_\nu^M = \partial_\mu \tilde{f}_{\mu\nu}$. An effective theory of this type reproduces the standard $\beta$-function for $g$ [19, 20]. Note that no term involving $f_{\mu\nu}B_{\mu\nu}$ is generated at this order.

Finally, one of the UV irrelevant vertices, $\mathcal{O}(1/\Lambda^2)$, has been singled out. This corresponds to the term governing the dynamics of the tensor field. In (4) it was included alongside the relevant terms in the first line. This vertex corresponds to the leading ghost free term in lowest order in a derivative expansion. It will be discussed at greater length presently.

The effective action (4) is in the spirit of [21] an appropriate initial condition at the scale $\Lambda$ for an ERG analysis of the low energy theory. Remember that it has been required for $\Lambda$ to be in the perturbative region of $SU(2)$ YM to ensure a reliable calculation of the coefficients in (4). In line with Abelian dominance it has also been required that $\Lambda >> M_{AD}$. There is another aspect to this last requirement. It can be seen that when an explicit mass $M_{AD}$ is introduced in the effective action the coefficients in (4) receive corrections in powers of $M_{\rm AD}^2/\Lambda^2$ and therefore it is necessary to have $\Lambda \gg M_{\rm AD}$ for reliability. Note that the scale $M_{\rm AD}$ is not to be confused with $\Lambda_{\rm QCD}$. This are distinct though expected to be related scales and $M_{\rm AD} > \Lambda_{\rm QCD}$.

**Exact renormalisation group and confinement.** Now it is shown that qualitatively the Abelian effective action (4) contains the relevant degrees of freedom for confinement. This is done within a functional analysis using ERG methods. The action is now truncated in two ways. First, the ghost sector is omitted as its inclusion would not change the present qualitative results. Second, only terms up to the quadratic order in a derivative expansion are retained, and therefore the kinetic term for the tensor field is kept.

The ERG equations are flow equations for an effective action $\Gamma_k$ with respect to the scale $k$. The $k$ dependence comes from IR cut-off functions $R_k$ explicitly inserted in the action [5]. The resulting flow equation has the structure of a one-loop equation

$$\partial_t \Gamma_k = \frac{1}{2} \operatorname{Tr} \left[ \partial_t R_k \left( \Gamma_k^{(2)} + R_k \right)^{-1} \right], \quad t = \ln k, \qquad (5)$$

where Tr stands for the sum over all fields and group indices as well as the integration over the coordinate space, whilst $\Gamma_k^{(2)}$ is the fully dressed 1PI functional. The effective action (4) suggests the following Ansatz,

$$\Gamma_k = \int_x \left\{ \frac{Z_A}{4} f_{\mu\nu}^2 + \frac{1}{4}(-Z_B \Box + M_B^2) B_{\mu\nu}^2 + \frac{Y}{2} \tilde{f}_{\mu\nu} B_{\mu\nu} + \frac{1}{2\xi} (\partial_\mu A_\mu)^2 \right\}, \qquad (6)$$

where $Z_A$, $Z_B$, $M_B^2$, $Y$ and $\xi$ are $k$ dependent functions. The use of this Ansatz reduces the complex equation (5) to a set of finite treatable coupled equations. The flow connects $\Gamma_\Lambda$ to $\Gamma_0$, the full effective action. Of course, a necessary condition is that the IR cut-off functions vanish as $k \to 0$.

Next, it is shown that this Ansatz contains enough information to exhibit a confining phase at low energies. A signature for confinement is sought in the form of a $1/p^4$ singularity in the gauge field propagator, a feature that has been shown to lead to an area law in a Wilson loop [22].

The initial condition for the ERG flow is assigned by taking $\Gamma_\Lambda \approx S_{\rm eff}$. For the new coefficient $Z_B$ this reduces to $Z_B(\Lambda) \approx 0$ which simply reflects the fact that $B_{\mu\nu}$ was introduced as an auxiliary field. Note also from (4) that $M_B^2(\Lambda) < 0$. Only a more complete treatment of the flow equations including the ghost sector can provide information on whether $M_B^2$ becomes positive at low energies. For the present purpose, it is sufficient to assume that near the confinement scale $M_B^2 > 0$.

From the Ansatz (6) the propagators can be expressed as functionals of the renormalisation functions. The gauge field propagator is

$$(P_{AA})_{\mu\nu} = \left(\delta_{\mu\nu} - \frac{p_\mu p_\nu}{p^2}\right) \frac{Z_B\, p^2 + M_B^2}{Z_A Z_B\, p^4 + (M_B^2 Z_A - Y^2)\, p^2} + \xi\, \frac{p_\mu p_\nu}{p^4}. \qquad (7)$$

As $\xi \to 0$ at low energies, the denominator of the propagator (7) is dominated by the $p^4$ term if

$$M_B^2(k)\, Z_A(k) = Y(k)^2. \qquad (8)$$

Then the gauge field propagator has a $1/p^4$ singularity when

$$Z_B(k_c)\, p^2 \ll M_B^2(k_c) \quad \Rightarrow \quad \sqrt{p^2} \ll \frac{M_B(k_c)}{\sqrt{Z_B(k_c)}} := \Lambda_{\text{QCD}}. \qquad (9)$$

It is expected that (8) is an IR stable quasi-fixed point as in the approximations considered in [7].

In a recent work, Ellwanger and Wschebor [23], have found that for an effective gauge theory where it is assumed that the charged fields as well as the ghosts have been integrated over, the equivalent to condition (8) is relaxed to an inequality which in the present case would read $M_B^2(k)\, Z_A(k) - Y(k)^2 < 0$, for $k > 0$. The equality is recovered in the IR limit $k \to 0$.

**Confinement and the dynamics of the $B_{\mu\nu}$ field.** The IR $1/p^4$ singularity occurs because: (a) there is a term proportional to $p^4$ in the denominator of the propagator that becomes prominent when the condition (8) holds; (b) at small momentum, when (9) is fulfilled, the numerator becomes $p^2$ independent. The proportionality factor to $p^4$ is $Z_A Z_B$. Clearly, for this factor to be non zero the $B_{\mu\nu}$ field must become dynamical at low energies. Therefore, somewhere in between the deep UV region $k \simeq \Lambda$ and the confinement scale $\Lambda_{\text{QCD}}$, the tensor field kinetic vertex must undergo a crossover that will make it relevant in the IR. This crossover scale is expected to be linked to the Abelian dominance scale $M_{\text{AD}}$. Above this scale the dynamics of $B_{\mu\nu}$ is protected by the still unsuppressed $\phi_\mu$ fields by $\mathcal{O}(1/\Lambda^2)$. However, below $M_{\text{AD}}$ the off-diagonal fields are suppressed and are counterbalanced by the Abelian tensor field. Consequently, $Z_B$ becomes equally relevant in the Abelian dominated regime when compared with the other renormalisation functions which makes the above scenario leading to a $1/p^4$ behaviour plausible. Polonyi's view of confinement as an irrelevant-to-relevant crossover [24] of some UV irrelevant operator is in line with the present scenario.

**Duality in the confinement phase.** In the confining regime when conditions (8) and (9) are satisfied the effective action reduces to

$$\Gamma_{\text{conf}} = \int_x \Big\{ \frac{1}{4Z_A}(Z_A f_{\mu\nu} + Y\tilde{B}_{\mu\nu})^2 + \frac{Z_B}{4}(\partial_\rho \tilde{B}_{\mu\nu})^2 \Big\}. \tag{10}$$

Now consider the transformation $(a_\mu, B_{\mu\nu}) \to (b_\mu, \theta)$ defined implicitly by

$$b_{\mu\nu} = \frac{\epsilon_{\mu\nu\rho\sigma}}{2\sqrt{Z_A}}(Z_A f_{\rho\sigma} + Y\tilde{B}_{\rho\sigma}), \ \partial_\sigma \theta + \sqrt{\frac{3Y^2}{Z_A Z_B}}\, b_\sigma = \sqrt{\frac{Z_B}{12}}\, \epsilon_{\mu\nu\rho\sigma}\, \partial_\mu \tilde{B}_{\nu\rho}, \tag{11}$$

where $b_{\mu\nu} = \partial_\mu b_\nu - \partial_\nu b_\mu$. In terms of the new variables the confining action (10) becomes

$$\tilde{\Gamma}_{\text{conf}} = \int_x \Big\{ \frac{1}{4} b_{\mu\nu}^2 + \frac{1}{2}\Big(\partial_\mu \theta + \sqrt{\frac{3Y^2}{Z_A Z_B}}\, b_\mu\Big)^2 \Big\}. \tag{12}$$

By substituting $\Phi = \rho\, e^{-i\theta}$ with $\rho$ frozen above, the confining action (12) is shown to describe a type-II dual superconductor with dual London penetration depth $\lambda_b \sim \sqrt{Z_A Z_B}/Y$. This action is analogue to the one found in [7].

The change of variables (11) is a dual transformation in the usual sense that the equations of motion for $(a_\mu, B_{\mu\nu})$ become Bianchi identities for $(b_\mu, \theta)$ and vice-versa. Note that the mismatch in degrees of freedom arises from an invariance in the transformation (11) with respect to the shift

$$a_\mu \to a_\mu + \epsilon_\mu, \ \tilde{B}_{\mu\nu} \to \tilde{B}_{\mu\nu} - \frac{Z_A}{Y}(\partial_\mu \epsilon_\nu - \partial_\nu \epsilon_\mu) \tag{13}$$

for arbitrary $\epsilon_\mu$. This type of shift is used to generate monopole configurations in Abelian theories on the lattice.

**Summary and discussion.** The analytic approach described here provides a promising framework to study the low energy confining regime of non-Abelian gauge theories. To bring it to a more quantitative level further research is necessary. ERG methods have been combined with the benefits of working in a MAG to study the monopole mechanism for confinement in $SU(2)$ YM. As Abelian dominance has been observed on the lattice with a MAG gauge fixing [4] it is expected that that an intermediate Abelian dominance scale $M_{\text{AD}} > \Lambda_{\text{QCD}}$ is dynamically generated. Then starting from a pure $SU(2)$ YM theory a perturbative Abelian effective action was derived. Because there is a scale separation in a MAG this action is suitable to be integrated using ERG equations. It is shown that under certain conditions confinement occurs if the ERG flow at low energies settles about an IR stable quasi-fixed point. Confinement hinges on an irrelevant-to-relevant crossover for the kinetic term of the tensor field and it is expected that the crossover takes place at the scale of Abelian dominance $M_{\text{AD}}$. Finally, it was shown that in the confining phase the effective action describes the physics of a type II magnetic Abelian Higgs system in the broken phase.

**Acknowledgements.** The author is grateful to Jan Pawlowski and Denjoe O'Connor for useful discussions and Štefan Olejník and Jeff Greensite for putting together such an interesting workshop.